\documentstyle[12pt,espcrc1,epsfig]{article}
\let\section=\subsection     \let\subsection=\subsubsection                
\begin{document}
\newcommand{\be}{\begin{eqnarray}}
\newcommand{\ee}{\end{eqnarray}}
\newcommand\del{\partial}
\newcommand\barr{|}
\newcommand\cita{\cite}
\newcommand\half{\frac 12}
\begin{center}
   {\large \bf CHIRAL SYMMETRY AND THE SPECTRUM}\\[2mm]
   {\large \bf OF THE QCD DIRAC OPERATOR}\\[5mm]
   J.J.M.~VERBAARSCHOT\\[5mm]
   {\small \it  Department of Physics\\
   SUNY at Stony Brook, Stony Brook, NY11790, USA \\[8mm] }
\end{center}

\begin{abstract}\noindent
According to the Banks-Casher formula the chiral order parameter is
directly related to the spectrum of the Dirac operator.
In this lecture, we will argue that some properties
of the Dirac spectrum are universal and can be obtained from a random matrix 
theory with the global symmetries of the QCD partition function. 
In particular, this is true for the spectrum near zero on the scale of
a typical level spacing. Alternatively, the chiral order parameter 
can be characterized by the zeros of the partition function. 
We will analyze such  zeros for a random matrix model
at nonzero chemical potential.
\end{abstract}

\section{Introduction}
Many phenomena in nuclear physics, as for example
the lightness of the pion mass and the absence of parity doublets, can be
explained by the assumption that chiral symmetry is spontaneously broken.
This assumption has been confirmed by numerous lattice QCD simulations
(for a review see \cite{DeTar,Ukawa}). However, these studies also
show that chiral symmetry \cite{detaru1} 
is restored at a critical temperature of
$T_c \approx 140\,\, MeV$. In spite of steady progress
\cite{barbour}, the situation at nonzero baryon number 
density is much less clear \cite{everybody}. It seems that the quenched 
approximation does not work \cite{everybody,stephanov}, 
and the phase of the fermion determinant makes
unquenched simulations virtually impossible. 

The order parameter of the chiral phase transition is the chiral condensate.
It is directly related to the spectral density of the Euclidean 
Dirac operator \cite{Banks-Casher}. 
One of the questions we wish to address
is to what extent the Dirac spectrum 
shows universal features which 
can be obtained from a Random Matrix Theory (RMT) with the 
global symmetries of the QCD partition function  (chiral 
Random Matrix Theory (chRMT)). 
As is well-known from the study of
complex systems \cite{bohigas}, 
only correlations on a scale set by the eigenvalue spacing are given by RMT.
Such correlations may be important 
in mesoscopic systems. Typical examples are 
a finite nucleus \cite{Hauser},  small metallic
particles \cite{Denton}, quantum dots and 
disordered wires (see \cite{Beenakker} for a review). In particular, 
universal conductance fluctuations have been understood in the framework of 
RMT \cite{meso}. In lattice QCD simulations, with a mesoscopic
number of degrees of freedom, 
we expect to observe similar phenomena. In particular, 
in the mesoscopic range of QCD, for box size $L$ given by \cite{LS}
$
1/\Lambda \ll L \ll 1/m_\pi,
$
we expect to obtain exact results from RMT ($\Lambda$ is a typical hadronic 
scale and $m_\pi$ is the pion mass).

In spite of its success in explaining mesoscopic phenomena, I wish
to emphasize that there is no universality on a macroscopic scale and 
that RMT cannot be used to obtain quantitative predictions at this scale.
(Note, however, \cite{Voiculescu,zeeblue,engel,janik}.). 
This does not imply that RMT cannot be useful for the study
of macroscopic phenomena. They have  widely been used as $schematic$ models
for disorder, e.g., Anderson localization 
\cite{Anderson} and the Gross-Witten model \cite{GW}.
Because of problems in simulating  QCD at a nonzero baryon number density,
chRMT is an ideal laboratory to address this problem. One important success
is the understanding of the nature of the quenched approximation
\cite{stephanov}. Below, we will discuss
the phase structure of chRMT by means of the distribution 
of Yang-Lee zeros \cite{frank}.
\section{The Chiral order parameter}
The order parameter of the chiral phase transition,
$\langle \bar \psi \psi \rangle$,
is nonzero only below the critical temperature. 
As was shown in \cite{Banks-Casher} 
$\langle \bar \psi \psi \rangle$ is directly related to the eigenvalue density
of the QCD Dirac operator per unit four-volume 
\be
\Sigma \equiv 
|\langle \bar \psi \psi \rangle| =\frac {\pi \langle {\rho(0)}\rangle}V.
\label{bankscasher}
\ee
It is elementary to derive this relation.
The Euclidean Dirac operator for 
gauge field configuration $A_\mu$ is given by
$
D = \gamma_\mu (\del_\mu + i A_\mu).
$
For Hermitean gamma matrices 
$D$ is anti-hermitean with purely imaginary eigenvalues,
$
D\phi_k = i\lambda_k \phi_k,
$
and  spectral density given by
$
\rho(\lambda) = \sum_k \delta (\lambda - \lambda_k).
$
Because $\{\gamma_5, D\} = 0$, 
nonzero eigenvalues occur in pairs $\pm \lambda_k$. 
In terms of the eigenvalues of $D$ the QCD partition function
for $N_f$ flavors of mass $m$ can then be written as
\be
Z(m) = \langle \prod_k(\lambda_k^2 + m^2)^{N_f} \rangle,
\label{part}
\ee
where the average $\langle \cdot \rangle$ 
is over all gauge field configurations weighted according to the Euclidean
action.

The chiral condensate follows immediately from
the partition function (\ref{part}),
\be
\langle \bar \psi \psi \rangle  = \frac 1{VN_f}\del_m \log Z(m) 
=\frac 1V \langle \sum_k \frac {2m}{\lambda_k^2 + m^2}\rangle.
\ee
If we express the sum as an integral over the spectral density,
and take the thermodynamic limit before the chiral limit so that we have many
eigenvalues less than $m$ we recover (\ref{bankscasher}) (Notice
the order of the limits.).

Another way to characterize the chiral condensate is via the zeros of
the partition function \cite{vink,barbourqed}. 
For a finite number of degrees of freedom  
$Z(m)$ can be factorized as
$
Z(m) \sim \prod_k(m-m_k).
$
The chiral condensate is then given by 
\be
\langle \bar \psi \psi \rangle =
\frac 1{VN_f} \del_m \log Z(m) = \frac 1V \sum_k \frac 1{m-m_k}.
\ee
In the chirally broken phase zeros are located on a segment of
the imaginary axis that includes $m= 0$.
In the thermodynamic limit they coalesce into a cut and 
the chiral condensate shows a discontinuity each time $m$ crosses this cut.
In the chirally symmetric phase,
we expect to find  a cut $away$ from the imaginary axis.

The eigenvalues of the Dirac operator 
$
D= \gamma_\mu (\del_\mu + i A_\mu) + \gamma_0 \,\mu
$
at nonzero chemical potential
are scattered in the complex plane (see for example \cite{everybody}). 
For a finite number of degrees of freedom the partition function, $Z(m,\mu)$, 
is a 
polynomial in $m$ and $\mu$.  The condensate can be related to
either to the spectral density or to the zeros of $Z(m,\mu)$. An 
alternative order parameter is the baryon density 
$
n_B =  \del_\mu \log Z(m,\mu)/{N_f V}.
$
We can differentiate with respect to $\mu$ before or after averaging over
the gauge fields. In the first case the baryon number density follows
from the spectral density of $\gamma_0 (D+ m)$ 
with eigenvalues scattered in the complex plane. 
In the second case the baryon number 
density follows from the zeros of $Z(m,\mu)$ in the complex
$\mu$ plane.
\section{The Dirac Spectrum}
An important consequence of the Bank-Casher formula (\ref{bankscasher})
is that the eigenvalues near zero virtuality are spaced as  
$
\Delta \lambda = 1/{\rho(0)} = {\pi}/{\Sigma V}.
$
This should be contrasted with the eigenvalue spectrum 
of the non-interacting 
Dirac operator. Then
$
\rho^{\rm free}(\lambda) \sim V\lambda^3
$
which leads to an eigenvalue spacing of $\Delta \lambda \sim 1/V^{1/4}$.
Clearly, the presence of gauge fields lead to a strong modification of
the spectrum near zero virtuality. Strong interactions result in the 
coupling of many degrees of freedom leading to extended states and correlated
eigenvalues.
On the other hand, for uncorrelated eigenvalues, the eigenvalue distribution
factorizes and we have $\rho(\lambda) \sim \lambda^{2N_f+1}$, i.e. no breaking
of chiral symmetry. 

Numerous studies have shown that spectral correlations of complex systems
on a scale set by the level spacing 
are universal, i.e. they do not depend on the dynamics
of the system and are completely determined by symmetries. 
Because the
QCD Dirac spectrum is symmetric about zero,
we have two different types of eigenvalue correlations:
correlations in the bulk of the spectrum and spectral correlations near zero 
virtuality. In the context of
chiral symmetry we wish to study 
the spectral density near zero virtuality. Because the
eigenvalues are spaced as $1/\Sigma V$ it is natural to introduce the
microscopic spectral density
\be
\rho_S(u) = \lim_{V\rightarrow \infty} \frac 1{V\Sigma}
\rho\left( \frac u{V\Sigma} \right).
\ee
The dependence on the macroscopic variable $\Sigma$  has been eliminated
and therefore $\rho_S(u)$ is a perfect candidate for a universal function.
\section{Spectral universality}
Spectra for a wide range of complex quantum systems 
have been studied both experimentally \cite{Haq,Guhr,Koch} and numerically
\cite{bohigas,selig,drozdz}. 
One basic observation
has been that the scale of variations of the average spectral
density and the scale of the spectral fluctuations separate. 
This allows us to unfold the spectrum, i.e. we rescale the 
spectrum in units of the local average level spacing. The fluctuations of the
unfolded spectrum can be measured by suitable statistics. We will consider the
nearest neighbor spacing distribution, $P(S)$, the
number variance, $\Sigma_2(n)$, and the $\Delta_3(n)$ statistic. The number
variance is defined as the variance of the number of levels in a stretch of
the spectrum that contains $n$ levels on average,
and $\Delta_3(n)$ is obtained from $\Sigma_2(n)$ by averaging over 
a smoothening kernel.  

These statistics can be obtained analytically for the invariant random matrix
ensembles (see \cite{bohigas,Mehta})
defined as ensembles of Hermitean 
matrices with independently distributed Gaussian matrix elements. 
Depending on the anti-unitary symmetry, the matrix elements are real, complex
or quaternion real. The corresponding Dyson index is given by
$\beta = 1,\, 2 $, and $4$, respectively. 
The nearest neighbor spacing distribution is given by 
$P(S) \sim S^{\beta}\exp(-a_\beta S^2)$. 
The asymptotic behavior of $\Sigma_2(n)$
and $\Delta_3(n)$ is
given by $\Sigma_2(n) \sim (2/\pi^2\beta) \log(n)$
and $\Delta_3(n) \sim \beta \Sigma_2(n)/2$. For uncorrelated eigenvalues
one finds that $P(S) = \exp(-S)$, $\Sigma_2(n) = n$ and $\Delta_3(n) = n/15$.
Characteristic features of random matrix correlations are
level repulsion at short distances and a strong suppression
of fluctuations at large distances.

Numerous studies have shown that 
the spectral correlations of a classically chaotic systems are given by RMT.
This conjecture has been strengthened by 
recent analytical arguments \cite{berry,andreev,kick} and universality
arguments  \cite{bz}.
 
\section{Chiral random matrix theory}
In this section we will introduce an instanton liquid inspired \cite{Shuryak}
RMT for the QCD partition function. 
In the spirit of the invariant random matrix ensembles 
we construct a model for the Dirac
operator with the global symmetries of the QCD partition function as input, but
otherwise  Gaussian random matrix elements. 
The chRMT that obeys these conditions is defined by
\cite{SVR,V,VZ}
\be
Z_\nu^\beta = \int DW \prod_{f= 1}^{N_f} \det({\rm \cal D} +m_f)
e^{-\frac{N\Sigma^2 \beta}4 {\rm Tr}W^\dagger W},\quad{\rm with}\quad
\label{zrandom}
{\cal D} = \left (\begin{array}{cc} 0 & iW\\
iW^\dagger & 0 \end{array} \right ),
\ee
and $W$ is a $n\times m$ matrix with $\nu = |n-m|$ and
$N= n+m$. The matrix elements of $W$ are either real ($\beta = 1$, chiral
Gaussian Orthogonal Ensemble (chGOE)), complex
($\beta = 2$, chiral Gaussian Unitary Ensemble (chGUE)),
or quaternion real ($\beta = 4$, chiral Gaussian Symplectic Ensemble (chGSE)).

This model reproduces the following symmetries of the QCD partition
function:
{\it i)} The $U_A(1)$ symmetry. All nonzero eigenvalues of the random matrix
Dirac operator occur in pairs $\pm \lambda$.
{\it ii)}  The topological structure of the QCD partition function. The 
Dirac matrix has exactly $|\nu|\equiv |n-m|$ zero eigenvalues. This identifies
$\nu$ as the topological sector of the model.
{\it iii)} The flavor symmetry is the same as in QCD \cite{SmV}. 
{\it iv)} The chiral symmetry is broken spontaneously with 
chiral condensate given by
$                                                      
\Sigma = \lim_{N\rightarrow \infty} {\pi \rho(0)}/N.
$
($N$ is interpreted as the (dimensionless) volume of space
time.)
{\it v)} The anti-unitary symmetries. 
For  fundamental fermions the matrix elements of the Dirac operator are complex
for $N_c \ge 3$ ($\beta= 2$) but can be chosen real for $N_c = 2$ ($\beta =1$).
For adjoint fermions they can be arranged 
into real quaternions ($\beta = 4$).

The ensemble of matrices in (\ref{zrandom})
is also known as the Laguerre ensemble. Note that its
spectral correlations in the bulk of the spectrum are given by the
invariant random matrix ensemble with the same value of $\beta$ 
\cite{laguerre}.
Both types of microscopic correlations are 
stable against deformations of the ensemble. 
This has been shown by a variety of different arguments
\cite{zee,sener,Damgaard,ambjorn}. 

Below we will discuss the microscopic spectral density. 
For $N_c = 3$, $N_f$ flavors and topological charge $\nu$
it is given by \cite{V}
\be
\rho_S(u) = \frac u2 \left ( J^2_{a}(u) -
J_{a+1}(u)J_{a-1}(u)\right),
\label{micro}
\ee
where $a = N_f + \nu$. The result for $N_c =2$, which is more complicated,
is given in \cite{V2}, and the result for the symplectic ensemble is
derived in \cite{nagao}.  

Together with the invariant random matrix ensembles, the chiral ensembles are
part of a larger classification scheme. As pointed out in \cite{class}, 
there
is a one to one correspondence between random matrix theories and symmetric
spaces.

\section{Lattice QCD results}
Recently, Kalkreuter \cite{Kalkreuter}
calculated $all$ eigenvalues of the lattice Dirac
operator both for Kogut-Susskind (KS) fermions and Wilson fermions
for lattices 
\begin{center}
\begin{figure}[!ht]
\centering\includegraphics[width=140mm,angle=0]
{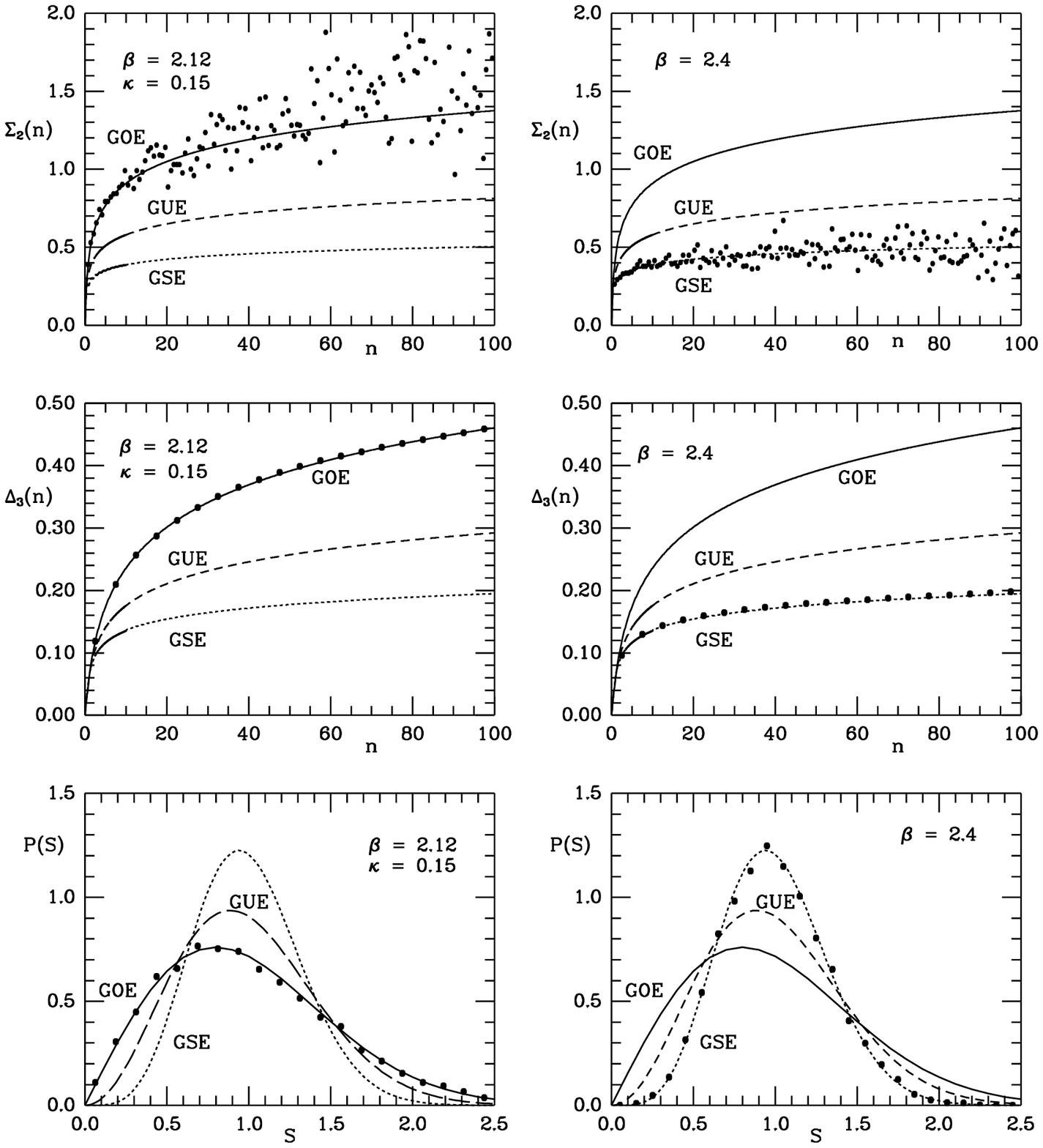}
\begin{center}
\begin{minipage}{13cm}
\baselineskip=12pt
\vspace*{0.3 cm}
{\begin{small}
Fig. 1. Spectral correlations of Dirac eigenvalues for Wilson fermions
(upper) and KS-fermions (lower). \end{small}}
\end{minipage}
\end{center}
\label{correlations}
\end{figure}
\end{center}
\noindent
as large as $12^4$. 
In the the case of $SU(2)$
the anti-unitary symmetry of the KS and the Wilson Dirac operator is
different  \cite{Teper,HV}. 
For KS fermions the
Dirac matrix can be arranged into real
quaternions, whereas 
the $Hermitean$ Wilson Dirac
matrix $\gamma_5 D^{\rm Wilson}$ can be chosen real.  
 Therefore, we expect that the eigenvalue
correlations are described by the GSE and the GOE, respectively \cite{HV}.
In Fig. 1 we show results for  $\Sigma_2(n)$,
$\Delta_3(n)$ and $P(S)$. 
The results for KS fermions are for 4 dynamical flavors
 with $ma = 0.05$ on a $12^4$ lattice. The results for Wilson fermion were
obtained for two dynamical flavors on a $8^3\times 12$ lattice.
Other statistics are discussed in \cite{HKV}.
\begin{center}
\begin{figure}[!ht]
\centering\includegraphics[width=80mm,angle=270]{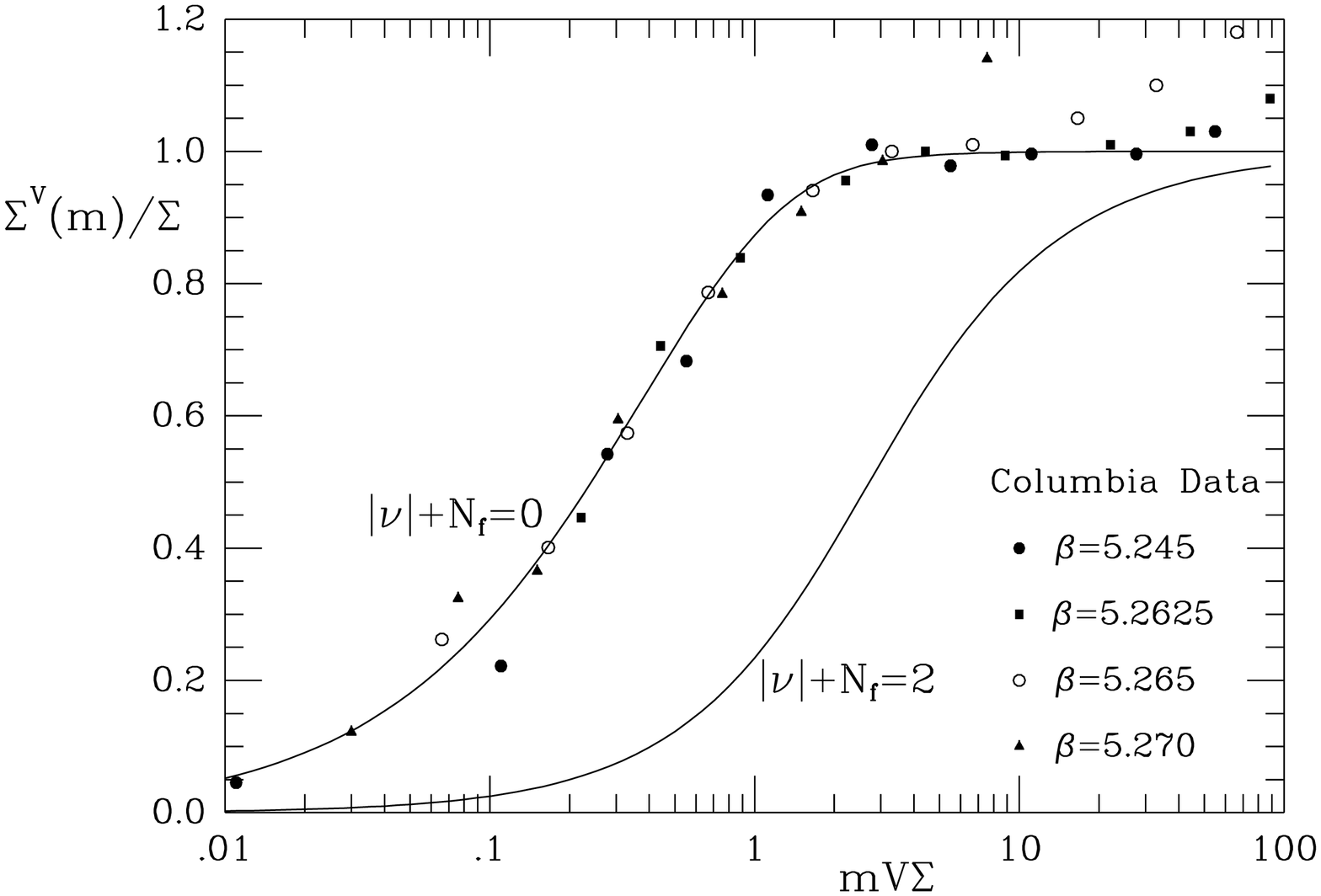}
\begin{center}
\begin{minipage}{13cm}
\baselineskip=12pt
\vspace*{ 0.3cm}
{\begin{small}
Fig. 2. The valence quark mass dependence of the chiral condensate.
\end{small}}
\end{minipage}
\end{center}
\label{columbia}
\end{figure}
\end{center}

Lattice studies of the microscopic spectral density are in progress 
and preliminary results are promising \cite{meyer}. However,
an alternative way to probe the Dirac spectrum was introduced by the
Columbia group \cite{Christ}. They studied the valence quark mass dependence
of the Dirac operator, i.e. $\Sigma(m) = \frac 1N \int d\lambda \rho(\lambda)
2m/(\lambda^2 +m^2)$, for a fixed sea quark mass. In the mesoscopic range,
the valence quark mass dependence can be obtained analytically from
the microscopic spectral density (\ref{micro}) \cite{vplb},
\be
\frac {\Sigma(x)}{\Sigma} = x(I_{a}(x)K_{a}(x)
+I_{a+1}(x)K_{a-1}(x)),
\label{val}
\ee
where $x = mV\Sigma$ is the rescaled mass and $a = N_f+\nu$.
In Fig. 2 we plot this ratio as a function of $x$ for
lattice data of two dynamical flavors with  mass $ma = 0.01$ and $N_c= 3$ on a
$16^3 \times 4$ lattice.  We observe
that the lattice data for different values of $\beta$ fall on a single curve.
Moreover, in the mesoscopic range 
this curve coincides with the random matrix prediction for $N_f = \nu = 0$.
Apparently, the zero modes are completely mixed with the much larger number of
nonzero modes. For eigenvalues much smaller than the sea quark mass, we expect
to see the $N_f = 0$ eigenvalue correlations.

\begin{figure}[!ht]
\begin{minipage}[b]{0.5\linewidth}
\centering\includegraphics[width=70mm]{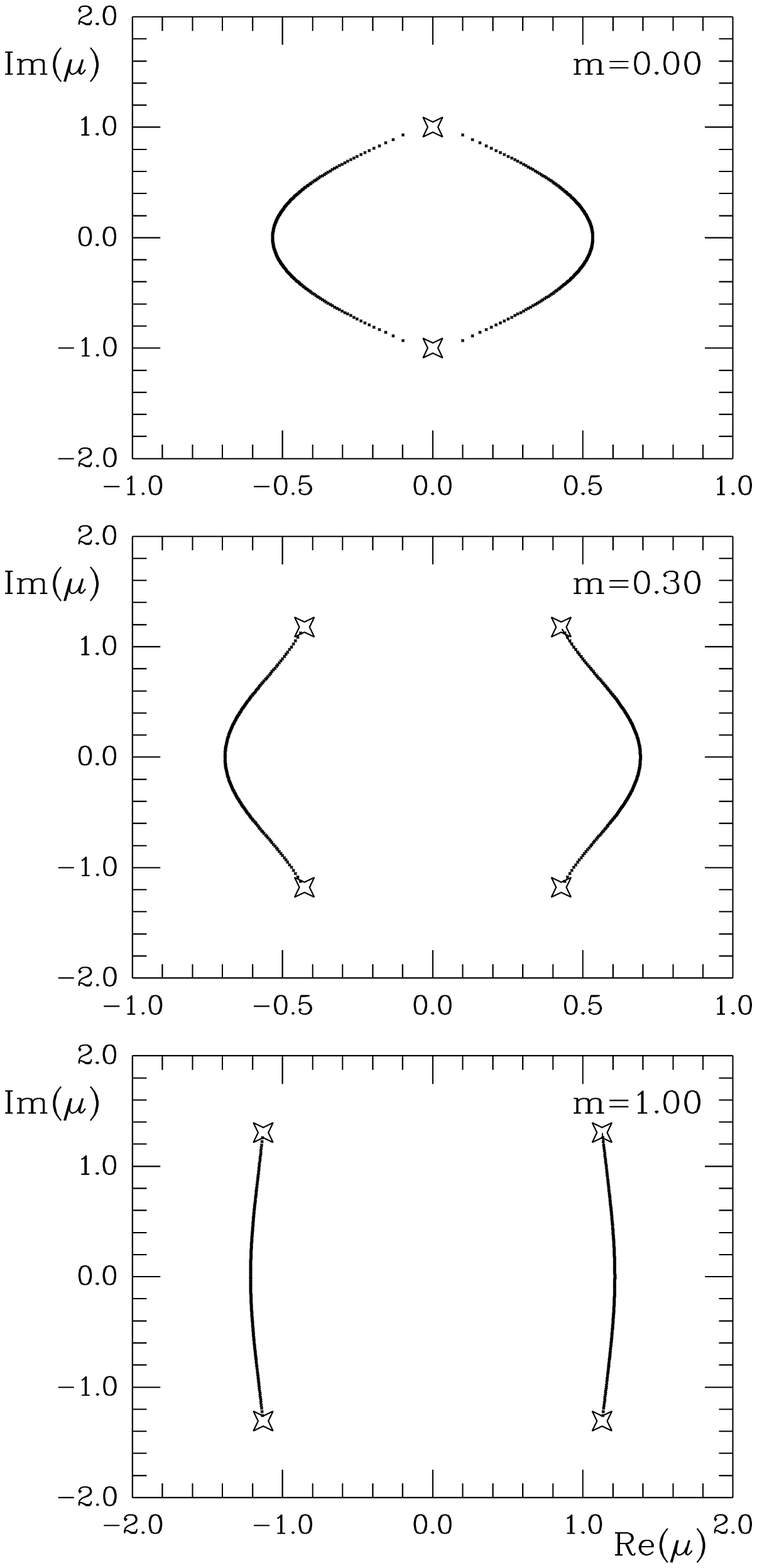}
\end{minipage}%
\begin{minipage}[b]{0.5\linewidth}
\centering\includegraphics[width=70mm]{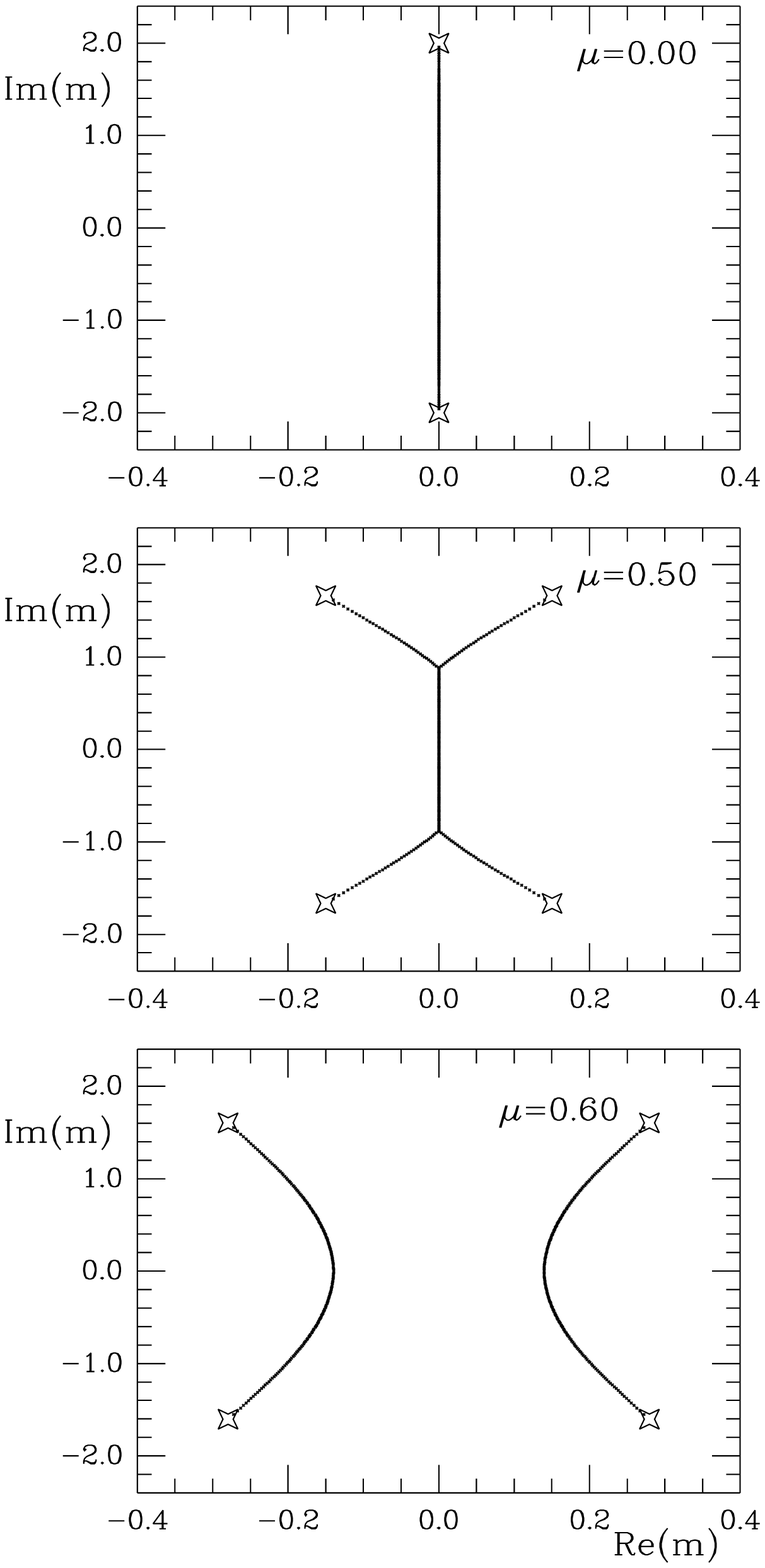}
\end{minipage}
\begin{center}
\begin{minipage}{13cm}
\baselineskip=12pt
\vspace*{0.3cm}
{\begin{small}
{Fig. 3. Zeros of the 
partition function in the complex $m$ and $\mu$ plane.}
\end{small}}
\end{minipage}
\end{center}
\label{lymu}
\end{figure}
\section{Chiral random matrix model at nonzero chemical potential}
At nonzero temperature and chemical potential the random matrix Dirac
operator in \ref{zrandom} is given by \cite{JV,Tilo,Stephanov1}
\be
{\cal D} = \left (\begin{array}{cc} 0 & iW +i\Omega_T +\mu \\
iW^\dagger+i\Omega_T + \mu & 0 \end{array} \right ),
\label{diracmatter}
\ee
where $\Omega_T= T \otimes_n(2n+1)\pi {\large\bf 1}$.

Inspired by \cite{Kocic},
the simplest model is obtained by keeping
only the lowest Matsubara frequency \cite{JV,Tilo}.
We wish to stress that  this model is a $schematic$ model of the QCD
partition function. Below, we will discuss a model with $\Omega_T$ 
absorbed by the random matrix and $\mu \ne 0$. Then the eigenvalues of
${\cal D}$ are scattered in the complex plane. In the quenched approximation
its distribution was obtained analytically \cite{stephanov} from 
the $N_f\rightarrow 0$ limit of a partition function with the determinant 
replaced by its absolute value. To this end the RMT
partition function was rewritten in terms a
$\sigma$-model amenable to a saddle point approximation. 
The $\sigma$-model
shows a second order phase transition at the boundary of the
spectrum  leading to a vanishing curvature and a diverging
two point function. This was confirmed by an explicit calculation
of this two-point function in \cite{janikmu}.

In the remainder of this section we consider the $unquenched$ 
partition function for one flavor. Using a multi-precision
package \cite{bailey}, we have calculated the (Yang-Lee)
zeros of the partition function in the complex $\mu$ and $m$ plane
for values of $n$ as large as 192. 
Results for $n=192$ are shown in Fig. 3. Notice that
the zeros fall on a  curve \cite{shrock}. 

From a saddle-point analysis it can be shown that for
zero mass the model shows a first order phase transition along the curve
$
{\rm Re}[ \mu^2 + \log(\mu^2) ] = -1
$  \cite{stephanov,halaszyl}. 
The discontinuity of $n_B$ requires that 
the zeros of the partition function fall along  this curve (see left
upper figure). At the endpoints (stars) two different solutions
of the saddle point equation (a cubic equation \cite{JV}) coalesce.
All other curves can be obtained from a saddle-point analysis as well.
A schematic picture of the phase structure in the
complex $m$ plane is shown in \cite{steph}. 

\section{Conclusions}
We have shown that the microscopic correlations of the QCD Dirac spectrum
can be explained by RMT and have obtained an analytical
understanding of the distribution of the eigenvalues closest to zero.
We have given an extension of this model to nonzero temperature
and chemical potential. Its phase structure has been mapped out unambiguously
by means of Yang-Lee zeros.

This work was partially supported by the US DOE grant
DE-FG-88ER40388. M.A. Halasz
is thanked for a critical reading of the manuscript.


\begin{thebibliography}{99}
\itemsep=0cm
\bibitem{DeTar}
C.~DeTar, {\it Quark-gluon plasma in numerical simulations of QCD}, in {\it
Quark gluon plasma 2}, R. Hwa ed., World Scientific 1995.
\bibitem{Ukawa}A.~Ukawa, 
Lattice 1996, hep-lat/9612011.
\bibitem{detaru1} C.~Bernard, T. Blum, C. DeTar, S. Gottlieb, U. Heller, 
J. Hetrick, K. Rummukainen, R. Sugar, D. Toussaint and M. Wingate,
Phys. Rev. Lett. {\bf 78} (1997) 598.
\bibitem{barbour}I. Barbour, S. Morrison and J. Kogut, 
hep-lat/9612012.
\bibitem{everybody}I. Barbour, N. Behihil, E. Dagotto, F. Karsch,
A. Moreo, M. Stone and H. Wyld, Nucl. Phys. {\bf B275} (1986) 296;
M. Lombardo, J. Kogut and D. Sinclair, hep-lat/9511026.
\bibitem{stephanov}M. Stephanov, Phys.\ Rev.\ Lett.\ {\bf 76} (1996) 4472.
\bibitem{Banks-Casher}T.~Banks and A.~Casher, Nucl. Phys. {\bf B169} (1980) 103.
\bibitem{bohigas}O.~Bohigas, M.~Giannoni, Lecture notes in Physics
{\bf 209} (1984) 1.
\bibitem{Hauser} W. Hauser and H. Feshbach, Phys. Rev. {\bf 87} (1952) 366;
J. Verbaarschot, H. Weidenm\"uller and M. Zirnbauer, Phys. Rep. 
{\bf 129} (1985) 367.
\bibitem{Denton}R. Denton, B. M\"uhlschlegel and D. Scalapino, Phys. Rev. 
Lett. {\bf 26} 1971.
\bibitem{Beenakker}C. Beenakker, Rev. Mod. Phys. (1997), 
cond-mat/9612179.
\bibitem{meso}Y. Imry, Europhysics Lett. {\bf 1} (1986) 249; 
S. Iida, H. Weidenm\"uller and J. Zuk, Phys. Rev. Lett. {\bf 64} (1990) 583;
Ann. Phys. (N.Y.) {\bf 200} (1990), 219.
\bibitem{LS}
H.~Leutwyler and A.~Smilga, Phys. Rev. {\bf D46} (1992) 5607.
\bibitem{Voiculescu}D. Voiculescu, K. Dykema and A. Nica, {\it Free Random 
Variables}, Am. Math. Soc., Providence RI, 1992.
\bibitem{zeeblue}A. Zee, Nucl. Phys. {\bf B474} (1996) 726.
\bibitem{engel}M. Engelhardt and S. Levit, hep-th/9609216.
\bibitem{janik}
R. Janik, M. Nowak, G. Papp, J. Wambach, and I. Zahed,
hep-ph/9609491.
\bibitem{Anderson}P. Anderson, Phys. Rev. {\bf 109} (1958) 1492.
\bibitem{GW}D. Gross and E. Witten, Phys. Rev. {\bf D21} (1980) 446;
S. Chandrasekharan, hep-th/9610225.
H. Sommers, A. Crisanti, H. Sompolinsky and Y. Stein, 
Phys. Rev. Lett. {\bf 60} (1988) 1895.
\bibitem{frank} C.N. Yang and T.D. Lee, Phys.\ Rev.\ {\bf 87} (1952) 104, 
410.
\bibitem{vink}J. Vink, Nucl.\ Phys.\ {\bf B323} (1989) 399.
\bibitem{barbourqed}
I. Barbour, A. Bell, M. Bernaschi, G. Salina and A. Vladikas, 
Nucl.\ Phys.\ {B386} (1992) 683.
\bibitem{Haq}R. Haq, A. Pandey and O. Bohigas,
Phys. Rev. Lett. {\bf 48} (1982) 1086.                                       
\bibitem{Guhr}
C. Ellegaard, T. Guhr, K. Lindemann, H.Q. Lorensen, J. Nygard
and M. Oxborrow, 
Phys. Rev. Lett. {\bf 75} (1995) 1546.
\bibitem{Koch}
S. Deus, P. Koch and L. Sirko, Phys. Rev. {\bf E 52} (1995) 1146;
H. Gr\"af, H. Harney, H. Lengeler, C. Lewenkopf, C. Rangacharyulu, A. Richter, 
P. Schardt and H. Weidenm\"uller, Phys. Rev. Lett. {\bf 69} (1992) 1296.
\bibitem{selig}T.~Seligman, J.~Verbaarschot, and M.~Zirnbauer,
Phys. Rev. Lett. {\bf 53}, 215 (1984);
T.~Seligman and J.~Verbaarschot, Phys. Lett. {\bf 108A} (1985) 183.
\bibitem{drozdz} S. Drozdz, A. Trellakis and 
J. Wambach, Phys. Rev. Lett. {\bf 76} (1996) 4891.
\bibitem{Mehta} F. Dyson and M. Mehta, J. Math. Phys. {\bf 4} (1963) 701.
\bibitem{berry}M. Berry, Proc. Roy. Soc. London {\bf A 400} (1985) 229.
\bibitem{andreev}
A. Andreev, O. Agam, B.D. Simons and B.L. Altshuler, 
cond-mat/9605204.
\bibitem{kick} A. Altland and M. Zirnbauer, Phys. Rev. Lett. {\bf 77} (1996)
4536.
\bibitem{bz}
A. Pandey, Ann. Phys. {\bf 134} (1981) 119; 
J. Ambjorn, J. Jurkiewicz and Y. Makeenko, Phys. Lett. {B251} (1990) 517;
E. Br\'ezin and A. Zee, Nucl. Phys. {\bf B402} (1993) 613;
C. Beenhakker, Nucl.Phys. {\bf B422} (1994) 515;
G. Hackenbroich and H. Weidenm\"uller, Phys. Rev. Lett. {\bf 74}
(1995) 4118;
S. Higuchi, C.Itoi, S.M. Nishigaki and N. Sakai, hep-th/9612237.
\bibitem{Shuryak}T. Sch\"afer and E. Shuryak, Rev. Mod. Phys. (1997), 
hep-ph/9610451.    
\bibitem{SVR}E. Shuryak and J. Verbaarschot,
Nucl. Phys. {\bf A560} (1993) 306.
\bibitem{V} J. Verbaarschot, Phys. Rev. Lett. {\bf 72} (1994) 2531; Phys. Lett.
{\bf B329} (1994) 351; Nucl. Phys. {\bf B427} (1994) 434.
\bibitem{VZ}J. Verbaarschot and I. Zahed,
Phys. Rev. Lett. {\bf 70} (1993) 3852.
\bibitem{SmV}
A. Smilga and J. Verbaarschot, Phys. Rev. {\bf D51} (1995) 829;
M. Halasz and J. Verbaarschot, Phys. Rev. {\bf D52} (1995) 2563.
\bibitem{laguerre}
D. Fox and P.Kahn, Phys. Rev. {\bf 134} (1964) B1152; 
(1965) 228; T. Nagao and M. Wadati, J. Phys. Soc. Jap. {\bf 60} (1991) 3298,
{\bf 61} (1992) 78, 1910.
\bibitem{zee}E. Br\'ezin, S. Hikami and A. Zee,
Nucl. Phys. {\bf B464} (1996) 411.
\bibitem{sener} A. Jackson, M. Sener and J. Verbaarschot, Nucl. Phys.
{\bf B479} [FS] (1996) 707.
\bibitem{Damgaard}S. Nishigaki, Phys. Lett. B (1996); G. Akemann, 
P. Damgaard, U. Magnea and S. Nishigaki, 
hep-th/9609174.
\bibitem{ambjorn}J. Ambjorn and G. Akeman, J. Phys. {A29} (1996) L555; Nucl. 
Phys. {\bf B482} (1996) 403.
\bibitem{V2}J. Verbaarschot, Nucl. Phys. {B426} (1994) 559.
\bibitem{nagao}T. Nagao and P.J. Forrester, Nucl. Phys. {\bf B435} (1995) 401.
\bibitem{class}F. Dyson, Comm. Math. Phys. {\bf 19} (1970) 235;
A. Altland, M. Zirnbauer, Phys. Rev. Lett. {\bf 76}
(1996) 3420;  M. Zirnbauer, J. Math. Phys. {\bf 37} (1996) 4986;
M.Caselle, 
cond-mat/9610017.
\bibitem{Kalkreuter}T. Kalkreuter,  Comp. Phys. Comm. {\bf 95} (1996) 1;
Phys. Lett. {\bf B276} (1992) 485; Phys. Rev. {\bf D48} (1993) 1926.
\bibitem{Teper}S. Hands and M. Teper, Nucl. Phys. {\bf B347} (1990)
819.
\bibitem{HV}M. Halasz and J. Verbaarschot,
Phys. Rev. Lett. {\bf 74} (1995) 3920.
\bibitem{HKV}M. Halasz, T. Kalkreuter and J. Verbaarschot, hep-lat/9607042,
Lattice 1996.
\bibitem{meyer} S. Meyer, private communication; T. Wettig, T. Guhr, A. 
Sch\"afer and H. Weidenm\"uller, hep-ph/9701387 (this proceedings).            
\bibitem{Christ}S. Chandrasekharan, Lattice 1994, 475;
S. Chandrasekharan and N. Christ, Lattice 1995, 527; N. Christ, Lattice 1996.
\bibitem{vplb}J. Verbaarschot, Phys. Lett. {\bf B368} (1996) 137.
\bibitem{JV}A. Jackson and J. Verbaarschot, Phys. Rev. {\bf D53} (1996)
7223.
\bibitem{Tilo}T. Wettig, A. Sch\"afer and H. Weidenm\"uller,
Phys. Lett. {\bf B367} (1996) 28.
\bibitem{Stephanov1}M. Stephanov, Phys. Lett. {\bf B275} (1996) 249.
\bibitem{Kocic}A. Kocic and J. Kogut, Nucl. Phys. {\bf B455} (1995) 229.
\bibitem{janikmu}
R. Janik, M. Nowak, G. Papp and I. Zahed,
Phys. Rev. Lett.{\bf 77} (1996) 4876. 
\bibitem{bailey}D. Bailey, NASA Ames RNR Technical Report RNR-94-013.
\bibitem{shrock} V. Matteev and R. Shrock, J.\ Phys.\ A: Math.\ Gen.\ {\bf 28}
(1995) 5235.
\bibitem{halaszyl}M. Halasz, A. Jackson and J. Verbaarschot, Phys. Lett.
{\bf B} (1997); in preparation.
\bibitem{steph}M. Stephanov, hep-lat/9607060, Lattice 1996.
\end{thebibliography}
\end{document}